\input harvmac
\input amssym
\input epsf

\Title{\vbox{\baselineskip12pt
\hbox{KUNS-1857}\hbox{\tt hep-th/0404121}}}
{\vbox{\centerline{Cosmological Constant Probing Shape Moduli}
	\vskip12pt\centerline{through Large Extra Dimensions}}}

\centerline{
Satoshi Matsuda\footnote{$^\dagger$}{\tt matsuda@yukawa.kyoto-u.ac.jp} and 
Shigenori Seki\footnote{$^\ddagger$}{\tt seki@gauge.scphys.kyoto-u.ac.jp}\footnote{$^*$}{Moved from 
Theoretical Physics Laboratory, The Institute of Physical and Chemical 
Research (RIKEN), Wako 351, Japan on April 1, 2004.}}

\bigskip

{\it \centerline{Department of Physics, 
Graduate School of Science}
\centerline{Kyoto University, Kyoto 606, Japan}}

\bigskip

\vskip .2in

\centerline{{\bf abstract}}

We consider compactification of extra dimensions 
and numerically calculate Casimir energy which is 
provided by the mass of Kaluza-Klein modes. 
For the extra space we consider a torus with shape moduli and 
show that the corresponding vacuum energy is represented as a function 
of the moduli parameter of the extra dimensions.
By assuming that the Casimir energy may be identified 
with cosmological constant, we evaluate the size of extra dimensions 
in terms of the recent data given by the Wilkinson Microwave 
Anisotropy Probe (WMAP) measurement and the supernovae observations. 
We suggest that the observed cosmological constant may probe 
the shape moduli of the extra space by the study of the Casimir energy 
of the compactified extra dimensions.

\vskip .1in

\Date{}

\lref\ADD{N.~Arkani-Hamed, S.~Dimopoulos and G.~Dvali, ``The Hierarchy Problem and New Dimensions at a Millimeter,''  Phys. Lett. B429 (1998) 263, {\tt hep-ph/9803315}.}
\lref\AADD{I.~Antoniadis, N.~Arkani-Hamed, S.~Dimopoulos and G.~Dvali, ``New Dimensions at a Millimeter to a Fermi and Superstrings at a TeV,'' Phys. Lett. B436 (1998) 257, {\tt hep-ph/9804398}.}
\lref\AWi{J.~Ambj\o rn and S.~Wolfram, ``Properties of the Vacuum 1. Mechanical and Thermodynamic,'' Annals Phys. 147 (1983) 1.}
\lref\AWii{J.~Ambj\o rn and S.~Wolfram, ``Properties of the Vacuum 2. Electrodynamic,'' Annals Phys. 147 (1983) 33.}
\lref\Di{K.~R.~Dienes, ``Shape versus Volume: Making Large Flat Extra Dimensions Invisible,'' Phys. Rev. Lett. 88 (2002) 011601, {\tt hep-ph/0108115}.}
\lref\Dii{K.~R.~Dienes, ``Beautified with Goodly Shape: Rethinking the Properties of Large Extra Dimensions,'' proceedings of the SUSY 2002 Conference, {\tt hep-ph/0211211}.}
\lref\DM{K.~R.~Dienes and A.~Mafi, ``Shadows of the Planck Scale: The Changing Face of Compactification Geometry,'' Phys. Rev. Lett. 88 (2002) 111602, {\tt hep-th/0111264}.}
\lref\FKMi{M.~Fukuma, Y.~Kono and A.~Miwa, ``A Mechanism of the Large-scale Damping in the CMB Anisotropy,'' {\tt hep-th/0312298}.}
\lref\FKMii{M.~Fukuma, Y.~Kono and A.~Miwa, ``Noncommutative Inflation and the Large-scale Damping in the CMB Anisotropy,'' {\tt hep-th/0401153}.}
\lref\GS{I.~M.~Gel'fand and G.~E.~Shilov, ``Generalized Functions,'' Volume 1 (Academic Press, 1964).}
\lref\HLZ{T.~Han, J.~D.~Lykken and R.-J.~Zhang, ``On Kaluza-Klein States from Large Extra Dimensions,'' Phys. Rev. D59 (1999) 105006, {\tt hep-ph/9811350}.}
\lref\MS{S.~Matsuda and S.~Seki, ``Proposal for Generic Size of Large Extra Dimensions,'' {\tt hep-ph/0307361}.}
\lref\RSi{L.~Randall and R.~Sundrum, ``A Large Mass Hierarchy from a Small Extra Dimension,'' Phys. Rev. Lett. 83 (1999) 3370, {\tt hep-ph/9905221}.}
\lref\RSii{L.~Randall and R.~Sundrum, ``An Alternative to Compactification,'' Phys. Rev. Lett. 83 (1999) 4690, {\tt hep-th/9906064}.}
\lref\WMAPi{C.~L.~Bennett {\it et al.}, ``First Year Wilkinson Microwave Anisotropy Probe (WMAP) Observations: Preliminary Maps and Basic Results,'' Astrophys. J. Suppl. 148 (2003) 1, {\tt astro-ph/0302207}; see also the Web page, {\tt http://map.gsfc.nasa.gov/}.}
\lref\WMAPii{D.~N.~Spergel {\it et al.}, ``First Year Wilkinson Microwave Anisotropy Probe (WMAP) Observations: Determination of Cosmological Parameters,'' Astrophys. J. Suppl. 148 (2003) 175, {\tt astro-ph/0302209}.}
\lref\Ro{M.~D.~Roberts, ``Vacuum Energy,'' {\tt hep-th/0012062}.}
\lref\Diii{K.~R.~Dienes, ``Solving the Hierarchy Problem without Supersymmetry or Extra Dimensions: An Alternative Approach,'' Nucl. Phys. B611 (2001) 146, {\tt hep-ph/0104274}.}
\lref\PP{E.~Ponton and E.~Poppitz, ``Casimir Energy and Radius Stabilization in Five and Six Dimensional Orbifolds,'' JHEP 0106 (2001) 019, {\tt hep-ph/0105021}.}
\lref\Ga{C.~L.~Gardner, ``Primordial Inflation and Present-Day Cosmological Constant from Extra Dimensions,'' Phys. Lett. B524 (2002) 21, {\tt hep-th/0105295}.}
\lref\Mi{K.~A.~Milton, ``Constraints on Extra Dimensions from Cosmological and Terrestrial Measurements,'' Grav. Cosmol. 8 (2002) 65, {\tt hep-th/0107241}.}
\lref\Ch{I.~O.~Cherednikov, ``On Casimir Energy Contribution to Observable Value of the Cosmological Constant,'' {\tt astro-ph/0111287}.}
\lref\Me{A.~C.~Melissinos, ``Vacuum Energy and the Cosmological Constant,'' {\tt hep-ph/0112266}.}
\lref\Gu{A.~Gupta, ``Contribution of Kaluza-Klein Modes to Vacuum Energy in Models with Large Extra Dimensions \& the Cosmological Constant,'' {\tt hep-th/0210069}.}
\lref\Mii{K.~A.~Milton, ``Dark Energy as Evidence for Extra Dimensions,'' Grav. Cosmol. 9 (2003) 66, {\tt hep-ph/0210170}.}
\lref\BLS{F.~Bauer, M.~Lindner and G.~Seidl, ``Casimir Energy in Deconstruction and the Cosmological Constant,'' {\tt hep-th/0309200}.}
\lref\El{E.~Elizalde, ``Cosmological Uses of Casimir Energy,'' {\tt hep-th/0311195}.}
\lref\ADM{N.~Arkani-Hamed, S.~Dimopoulos and J.~March-Russell, ``Stabilization of Sub-millimeter Dimensions: The New Guise of the Hierarchy Problem,'' Phys. Rev. D63 (2001) 064020, {\tt hep-th/9809124}.}
\lref\KMST{N.~Kaloper, J.~March-Russell, G.~D.~Starkman and M.~Trodden, ``Compact Hyperbolic Extra Dimensions: Branes, Kaluza-Klein Modes and Cosmology,'' Phys. Rev. Lett. 85 (2000) 928, {\tt hep-ph/0002001}.}
\lref\Ko{C.~Kokorelis, ``Exact Standard Model Structures from Intersecting D5-Branes,'' Nucl. Phys. B677 (2004) 115, {\tt hep-th/0207234}.}
\lref\CIM{D.~Cremades, L.~E.~Ibanez and F.~Marchesano, ``Standard Model at Intersecting D5-branes: Lowering the String Scale,'' Nucl. Phys. B643 (2002) 93, {\tt hep-th/0205074}.}
\lref\NOZ{S.~Nojiri, S.~D.~Odintsov and S.~Zerbini, ``Bulk versus Boundary (Gravitational Casimir) Effects in Quantum Creation of Inflationary Brane-world Universe,'' Class. Quant. Grav. 17 (2000) 4855, {\tt hep-th/0006115}.}
\lref\CENOZ{G.~Cognola, E.~Elizalde, S.~Nojiri, S.~D.~Odintsov and S.~Zerbini, ``Multi-graviton Theory from a Discretized RS Brane-world and the Induced Cosmological Constant,'' {\tt hep-th/0312269}.}
\lref\BK{M.~Byrne and C.~Kolda, ``Quintessence and Varying $\alpha$ from Shape Moduli,'' {\tt hep-ph/0402075}.}
\lref\PePo{M.~Peloso and E.~Poppitz, ``Quintessence from Shape Moduli,'' Phys. Rev. D68 (2003) 125009, {\tt hep-ph/0307379}.}
\lref\Rietal{A.~G.~Riess {\it et al.}, ``Observational Evidence from Supernovae for an Accelerating Universe and a Cosmological Constant,'' Astron. J. 116 (1998) 1009, {\tt astro-ph/9805201}.}
\lref\Peetal{S.~Perlmutter {\it et al.}, ``Measurements of Omega and Lambda from 42 High-Redshift Supernovae,'' Astrophys. J. 517 (1999) 565, {\tt astro-ph/9812133}.}
\lref\AC{T.~Appelquist and A.~Chodos, ``Quantum Dynamics of Kaluza-Klein Theories,'' Phys. Rev. D28 (1983) 772.}

\newsec{Introduction}

Extra dimensions have been introduced 
in order to solve the hierarchy problem \ADD, 
and a lot of related works including the milestone papers 
\refs{\ADD\AADD\RSi{--}\RSii} and later 
developments \refs{\ADM\NOZ\CIM\Ko{--}\CENOZ} have been done.
In these approaches, gauge fields are localized on a 
four dimensional brane, which is regarded as our real world, and 
only a graviton can propagate in the extra space, which is 
transverse to the brane. 
When the extra space is compact, there exist the Kaluza-Klein modes of 
the graviton. 
They contribute to Casimir energy on the four dimensional brane. 

On the other hand, the origin of cosmological constant is 
also a standing problem. 
Recently the supernovae \refs{\Rietal,\Peetal} 
and WMAP \refs{\WMAPi,\WMAPii} observations 
have revealed some parameters of our universe. 
These new results encourage not only the researchers of cosmology but also 
the particle physicists \refs{\FKMi\FKMii{--}\MS} 
toward challenging the resolution of the unsolved problems again. 
If we assume that the Casimir energy, or in other words the vacuum energy, 
can be regarded as the cosmological constant, 
we can estimate the size of extra dimensions in terms of the observational 
value of the cosmological constant \MS. 

When we calculate the vacuum energy, we have to perform a summation over
the infinite number of Kaluza-Klein modes. 
In this paper, we introduce a new mathematical technique to achieve this 
infinite summation, which provides 
 a reasonable regularization and 
makes it possible to calculate numerically the vacuum energy 
of all Kaluza-Klein modes. 

In \refs{\ADD,\AADD}, it is shown that the number of extra dimensions 
should be two from the consistency with the hierarchy problem. 
The choice of the two extra dimensions are also favourable 
from the viewpoint of the cosmological constant problem \MS.
So we consider the extra space of two dimensions in this paper. 
In particular we concentrate on a torus with shape moduli. 
The compactification on the torus and the Kaluza-Klein masses 
under a standard mass operator are analysed in \refs{\KMST\Di\DM\Dii\PP{--}\BK}.
In this paper we additionally impose $(\Bbb{Z}_2)^2$ symmetry 
or Dirichlet conditions on the torus. 
Since the standard mass operator is inconsistent with these boundary 
conditions, 
we propose a new mass operator incorporating the shape moduli  
in consistency with the boundary conditions on the torus. 
Using this operator, we calculate
the vacuum energy. 
The resulting expression depends on the shape moduli parameter, 
as a function of which the structure of the extra space can be probed 
by confronting the numerical values of the Casimir energy with 
the observed cosmological constant \refs{\Rietal\Peetal\WMAPi{--}\WMAPii}.

The paper is organized as follows.
In the section 2, we consider the $(\Bbb{Z}_2)^2$ symmetry 
and Dirichlet conditions on the torus with the shape moduli. 
Requiring the invariance under these conditions, 
we introduce the new mass operator with the shape moduli. 
In the section 3, we show the technical details of the regularization of 
the vacuum energy to which the infinite number of Kaluza-Klein modes 
contribute. 
We also present the numerical results of the calculation of the vacuum 
energy.
In the section 4, we identify these numerical values of the vacuum energy 
with the cosmological constant which is given 
by \refs{\Rietal\Peetal\WMAPi{--}\WMAPii} and 
evaluate the size of extra dimensions. 
The section 5 is devoted to conclusions and discussion.

\newsec{Shape Moduli and Mass Operator}\seclab\shapemod

Though various kinds of extra space are available, 
we consider a two dimensional torus, $T^2$, in this paper. 
In the covering space of $T^2$, 
we require in the standard way the two periodic conditions,
\eqn\toruscond{\eqalign{
(y_1,y_2) &\sim (y_1 + 2\pi R_1, y_2) , \cr
(y_1,y_2) &\sim (y_1 + 2\pi R_2 \cos \theta, y_2 + 2\pi R_2 \sin \theta).
}}
$R_1$ and $R_2$ denote the radii of two cycles, while 
$\theta$ parametrizes the shape moduli of $T^2$. 
After the Kaluze-Klein compactification on such a torus, 
the eigen functions for the usual $({\rm mass})^2$ operator 
$-\left[(\partial / \partial y_1)^2 + (\partial / \partial y_2)^2\right]$ 
are proportional 
to
\eqn\efI{
\exp\left[i{n_1 \over R_1}\left(y_1 - {y_2 \over \tan \theta}\right) 
+i{n_2 \over R_2}{y_2 \over \sin \theta}\right],
}
where $n_1, n_2 \in \Bbb{Z}$, 
and the eigen values \refs{\Di\DM{--}\Dii} become 
$n_1^2/R_1^2 + n_2^2/R_2^2 - 2n_1n_2 \cos\theta /(R_1R_2)$.
For these mass eigen values, we can numerically calculate, 
in the same way as in \MS,  the vacuum energy 
in which we have to sum up all the Kaluza-Klein modes.

Now we introduce new coordinates $(X_1,X_2)$ defined by
\eqn\newcoord{
X_1 \equiv {y_1 \over R_1} - {y_2 \over R_1 \tan \theta}, 
\quad X_2 \equiv {y_2 \over R_2 \sin \theta} .
}
The periodic condition \toruscond\ is rewritten as 
\eqn\toruscondII{\eqalign{
(X_1,X_2) &\sim (X_1 + 2\pi, X_2),\cr
(X_1,X_2) &\sim (X_1, X_2 + 2\pi).
}}
The eigen functions \efI\ are equal to $\exp (in_1X_1 + in_2X_2)$ 
with $n_i \in \Bbb{Z}\ (i=1,2)$. 

In addition, let us impose $(\Bbb{Z}_2)^2$ symmetry on $T^2$.
\bigskip
\vbox{\centerline{\epsfbox{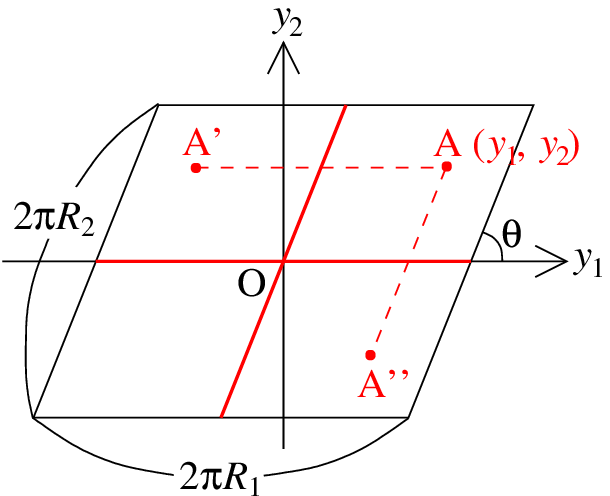}}}\nobreak
\medskip\nobreak
\centerline{\fig\figzIItorus{A fundamental region of $T^2$ 
in the covering space.}: A fundamental region of $T^2$ 
in the covering space.}
\bigskip\noindent
This symmetry is realized by the two mirror symmetry 
along the two thick lines 
represented in \figzIItorus. 
In  terms of the coordinates $(y_1,y_2)$, the symmetry is denoted 
by the following identification 
\eqn\zIIsym{\eqalign{
(y_1, y_2) &\sim \left(-y_1+{2y_2 \over \tan\theta}, y_2\right) , \cr
(y_1, y_2) &\sim \left(y_1 - {2y_2 \over \tan\theta}, -y_2\right) . 
}}
In \figzIItorus, the point 
A'$\left(-y_1+{2y_2 \over \tan\theta}, y_2\right)$ is 
identified with the point A$(y_1,y_2)$ by one $\Bbb{Z}_2$ symmetry and 
the point A''$\left(y_1 - {2y_2 \over \tan\theta}, -y_2\right)$ is also 
identified with the point A by the other $\Bbb{Z}_2$ symmetry.
Since the $(\Bbb{Z}_2)^2$ symmetry \zIIsym\ can be interpreted as 
the reflection symmetry given by  
\eqn\zIIsymII{
(X_1, X_2) \sim (-X_1, X_2), \quad (X_1, X_2) \sim (X_1, -X_2), 
}
the wave function under the $(\Bbb{Z}_2)^2$ symmetry, 
through the use of the coordinates $(X_1, X_2)$, 
should be an
even function with respect to each variable, $X_1$ or $X_2$. 
By the analogy of \efI, we may suppose that 
the wave function satisfying \toruscond\ and \zIIsym\ is
$\cos (n_1X_1)\cos (n_2X_2)$. 
But this function is not an eigen function of the $({\rm mass})^2$ operator 
$-\left[(\partial / \partial y_1)^2 + (\partial / \partial y_2)^2\right]$. 

So we propose a new mass operator 
which has $\cos (n_1X_1)\cos (n_2X_2)$ as an eigen function. 
First we note that the following differential operators simply 
change signs under the operation of $(\Bbb{Z}_2)^2$ symmetry on $T^2$, 
\zIIsym:
\eqn\diffop{
{\partial\over {\partial y_1}}={1\over R_1}{\partial\over{\partial X_1}}, 
\qquad
{\partial\over {\partial y_2}}+
{1\over{\tan\theta}}{\partial\over {\partial y_1}}=
{1\over {R_2\sin\theta}}{\partial\over{\partial X_2}}.
}
Though there are a variety of choices for the mass operator, 
the above observation has led us to consider, as the simplest toy model, 
the following generic $({\rm mass})^2$ operator defined by
\eqn\newmassope{-\left[
\left({1 \over R_1}{\partial \over \partial X_1}\right)^2
+ \left({1 \over R_2 \sin\theta}{\partial \over \partial X_2}\right)^2
\right]
=-\left[
{1 \over \sin^2\theta}{\partial^2 \over \partial y_1^2} 
+ {\partial^2 \over \partial y_2^2}
+ {2 \over \tan \theta}{\partial \over \partial y_1}{\partial \over 
\partial y_2}
\right]. 
}
The proposed $({\rm mass})^2$ operator \newmassope\ 
is invariant under the $(\Bbb{Z}_2)^2$ symmetry \zIIsym, and 
reduces to the standard one for 
$\theta=\pi/2$, that is, when the fundamental region of $T^2$ is 
rectangular.

In other words, though we can make a variety of choices for the metric 
in the extra space, the $(\Bbb{Z}_2)^2$ symmetry on $T^2$ has guided us 
to choose the simplest case of \newmassope\  for the mass operator. 
The corresponding metric of $T^2$ can be proved to be
$$
ds^2 =  dy_1^2 
-{1 \over \tan\theta}(dy_1dy_2 + dy_2dy_1) + 
{1 \over \sin^2\theta}dy_2^2 .
$$
This metric is naturally invariant under the $(\Bbb{Z}_2)^2$ symmetry \zIIsym, 
and again reduces to the standard rectangular one when $\theta=\pi/2$.

In the next section, we use the mass operator \newmassope\ and 
calculate the vacuum energy.

\newsec{Vacuum Energy and Boundary Conditions}

We consider the torus compactification with shape moduli. 
The torus  $T^2$ without shape moduli is the same as $(S^1)^{\otimes 2}$.  
The vacuum (Casimir) energy which is enhanced by the Kaluza-Klein modes 
derived from the compactification on the $d$ dimensional extra space 
$(S^1)^{\otimes d}$ has already been discussed in \MS.
In this section we consider 
the compactification on $T^2$ with shape moduli and the generic  
mass operator \newmassope\  as has been discussed 
in the section \shapemod. 
We then calculate the vacuum energy in our four dimensional space 
to which all the Kaluza-Klein modes from the extra space contribute. 

For simplicity we set $R_1 = R_2 \equiv R$.

\subsec{without a boundary condition}

The eigen functions compactified on $T^2$ with the mass operator \newmassope\ 
are described by $\exp(in_1X_1)\exp(in_2X_2)$ from \toruscondII, and 
the masses of Kaluza-Klein modes then become
$$
M = {1 \over R} \sqrt{n_1^2 + {n_2^2 \over \sin^2 \theta}}
\equiv {1 \over R} f_{n_1,n_2}(\theta) , \quad
n_1,n_2 \in \Bbb{Z} .
$$

We consider the zero point energies of these Kaluze-Klein modes and 
the vacuum energy of our four dimensional world is represented by
\eqn\vacene{
E = {1 \over 2} \sum_{n_1,n_2=-\infty}^\infty 
\int {d^3k \over (2\pi)^3}\sqrt{{\bf k}^2 + M^2} .
}
Note that, if the Kaluza-Klein modes are for gravitons, we 
have to insert 
a factor corresponding to 
the number of polarization degrees of freedom, 
$(2+2)(3+2)/2 -1=9$, in front of the right hand side of \vacene\ \HLZ.
In the following we consider only a scalar degree of freedom 
for simplicity.
Since the integration in \vacene\ is generally divergent, 
we use dimensional regularization for it and obtain
\eqn\regene{
E= -{1 \over 64\pi^2} \sum_{n_1,n_2=-\infty}^\infty 
M^4\left[\ln\left({\lambda^2 \over M^2}\right) + {3 \over 2}\right] ,
}
where the parameter $\lambda$ has a mass dimension.
Since the infinite summation for 
$\left(\sqrt{n_1^2 + n_2^2 /\sin^2\theta}\right)^4$ 
is calculated as
$$\eqalign{
\sum_{n_1,n_2=-\infty}^\infty f_{n_1,n_2}(\theta)^4
&= \sum_{n,m=-\infty}^\infty \left(1+ {1 \over \sin^4\theta}\right)n^4
+  \sum_{n,m=-\infty}^\infty {2 \over \sin^3\theta}n^2m^2 \cr
&= 4\left(1+ {1 \over \sin^4\theta}\right)\zeta(0)\zeta(-4)
+ {8 \over \sin^2\theta}\zeta(-2)^2 \cr
&= 0 , 
}$$
where $\zeta(-2) = \zeta(-4) = 0$, 
the vacuum energy \regene\ is reduced to 
\eqnn\casimir
$$\eqalignno{
E &= {1 \over 32\pi^2 R^4} \sum_{n_1,n_2=-\infty}^\infty 
{f_{n_1,n_2}}^4 \ln f_{n_1,n_2} \cr
&= {1 \over 32\pi^2 R^4} \sum_{n_1,n_2=-\infty}^\infty 
{d \over dx}{f_{n_1,n_2}}^x \biggr|_{x=4} . &\casimir
}$$
In calculating the vacuum energy, the infinite summation shows up  
and appears to diverge 
because of the contribution of the infinite number of Kaluza-Klein modes. 
It is known that in some special cases \refs{\AWi,\AWii} 
the vacuum energy can be 
regularized and are calculated exactly in terms of Epstein's zeta function. 
Since in general we can not use such a regularization, we suggest 
a new numerical method as follows. 
By the use of the Fourier transformation \GS, 
\eqn\fouriertrf{
\left(\sqrt{\sum_{i=1}^d \sigma_i^2}\right)^x = 
-{2^x\sin{\pi x \over 2} \over \pi^{{d \over 2}+1}}
\Gamma\left(1 + {x \over 2}\right)\Gamma\left({x+d \over 2}\right)
\int^\infty_{-\infty} {\exp\left(i\sum_{i=1}^d \sigma_iq_i\right) \over 
\left(\sqrt{\sum_{i=1}^d q_i^2}\right)^{x+d}} dq_1\cdots dq_d, 
}
which is valid for any real values of 
$(\sigma_1, \sigma_2, \cdots, \sigma_d)$, 
we obtain
\eqnn\frier
$$\eqalignno{
\sum_{n_1,n_2=-\infty}^\infty {f_{n_1,n_2}}^x
&= \sum_{n_1,n_2=-\infty}^\infty -{2^x \sin{\pi x \over 2} \over \pi^2}
\Gamma\left(1+{x \over 2}\right)^2
\int^\infty_{-\infty} {\exp (in_1q_1 + i{n_2 \over \sin\theta}q_2) \over 
\left(\sqrt{q_1^2 + q_2^2}\right)^{x+2}}dq_1dq_2 \cr
&= -{\sin{\pi x \over 2} \over \pi^{2+x}}
\Gamma\left(1+{x \over 2}\right)^2
\sin\theta \sum_{m_1,m_2=-\infty \atop (m_1,m_2)\neq(0,0)}^\infty
{1 \over (m_1^2+m_2^2\sin^2\theta)^{x+2 \over 2}}, &\frier
}$$
where we have used $\sum_{n=-\infty}^\infty \exp(inq) = 
2\pi \sum_{m=-\infty}^\infty \delta(q-2\pi m)$. 
The contribution of $(m_1,m_2) = (0,0)$ mode vanishes, 
which can be proven by employing a generalized function technique \GS.
From \casimir\ and \frier, the vacuum energy becomes 
\eqn\vacnobdr{
E = -{1 \over 64\pi^7 R^4}\Gamma(3)^2\sin\theta 
\sum_{m_1,m_2=-\infty \atop (m_1,m_2)\neq(0,0)}^\infty
{1 \over (m_1^2+m_2^2\sin^2\theta)^3}.
}

Since \vacnobdr\ is convergent, next we numerically calculate it. 
\bigskip
\vbox{\centerline{\epsfbox{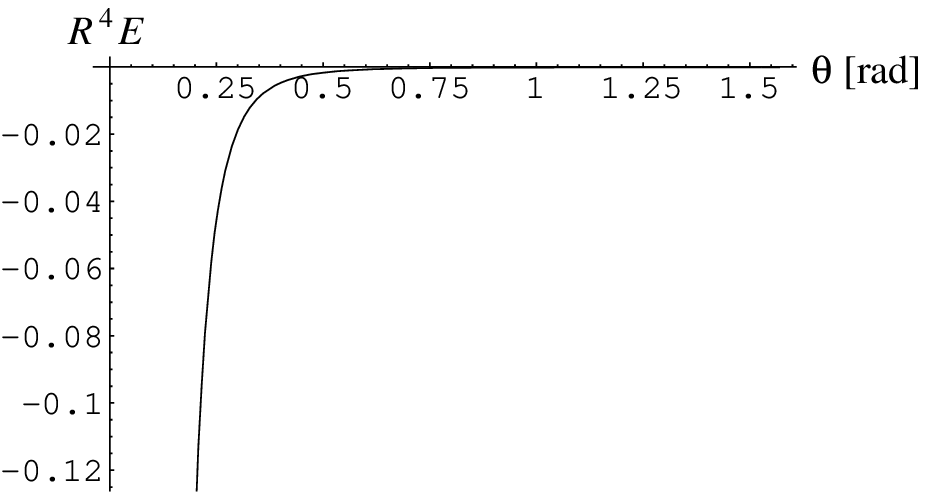}}}\nobreak
\medskip\nobreak
\centerline{
\fig\figvaceneNB{The $\theta$ dependence of the vacuum energy 
with no boundary condition. The values are negative for all $\theta$.}: 
The $\theta$ dependence of the vacuum energy 
with no boundary condition. }
\nobreak
\vskip -1mm
\centerline{
The values are negative for all $\theta$.
}
\bigskip\noindent
The numerical value of $R^4E$ which depends on $\theta$ is 
represented in \figvaceneNB. 
$R^4E$ is negative definite for all $\theta$.

\subsec{with $(\Bbb{Z}_2)^2$ symmetry}

We consider $(\Bbb{Z}_2)^2$ symmetry on $T^2$. 
Since the wave function should be an even function of $X_1$ and $X_2$ 
respectively  
on account of the condition \zIIsymII,  
the eigen function for the mass operator \newmassope\ becomes proportional to
$$
\cos (n_1 X_1) \cos (n_2 X_2) 
= \cos n_1\left({y_1 \over R_1} - {y_2 \over R_1\tan \theta}\right) 
\cos n_2 \left({y_2 \over R_2 \sin\theta}\right), 
\quad n_1,n_2 \in \{0,\Bbb{N}\}.
$$
We then obtain the masses of Kaluza-Klein modes, 
\eqn\kkmasszII{
M = {1 \over R}f_{n_1,n_2}(\theta), 
\quad n_1,n_2 \in \{0,\Bbb{N}\} .
}
Note that, 
since the eigen functions consist of cosine functions, $n_1$ and $n_2$ are 
zero or positive integers. 
In terms of \kkmasszII, we analyse the vacuum energy, and obtain 
\eqnn\vacEzIIi
$$\eqalignno{
E_{\Bbb{Z}_2} &= -{1 \over 64\pi^2} \sum_{n_1,n_2=0}^\infty 
M^4\left[\ln {\lambda^2 \over M^2} + {3 \over 2} \right] \cr
&= {1 \over 32\pi^2R^4}\sum_{n_1,n_2=0}^\infty {f_{n_1,n_2}}^4 
\ln f_{n_1,n_2} \cr
&= {1 \over 32\pi^2R^4}
\left[{1 \over 2}\sum_{n_1=-\infty}^\infty\left(1 + \delta_{n_1,0}\right)\right]
\left[{1 \over 2}\sum_{n_2=-\infty}^\infty\left(1 + \delta_{n_2,0}\right)\right]
{f_{n_1,n_2}}^4 \ln f_{n_1,n_2} \cr
&= {1 \over 128\pi^2R^4}\Biggl[ \sum_{n_1,n_2=-\infty}^\infty 
{f_{n_1,n_2}}^4\ln f_{n_1,n_2} \cr
&\phantom{= {1 \over 128\pi^2R^4}\Biggl[ } 
+ \sum_{n_1=-\infty}^\infty {f_{n_1,0}}^4 \ln f_{n_1,0}
+ \sum_{n_2=-\infty}^\infty {f_{0,n_2}}^4 \ln f_{0,n_2}
\Biggr] . &\vacEzIIi
}$$
where the last term has been dropped due to the identity  
${f_{0,0}}^4 \ln f_{0,0}=0$. 
The first term in \vacEzIIi\ has already been considered
in the previous subsection,  
and from \frier\ we get
\eqnn\infsumnn
$$\eqalignno{
\sum_{n_1,n_2=-\infty}^\infty {f_{n_1,n_2}}^4\ln f_{n_1,n_2} 
=& \sum_{n_1,n_2=-\infty}^\infty {d \over dx} {f_{n_1,n_2}}^x \biggr|_{x=4}\cr
=& -{1 \over 2\pi^5}\Gamma(3)^2 \sin\theta 
\sum_{m_1,m_2=-\infty \atop (m_1,m_2)\neq(0,0)}^\infty 
{1 \over \left(m_1^2+m_2^2\sin^2\theta\right)^3}. &\infsumnn
}$$
Using \fouriertrf, we calculate the second term in \vacEzIIi\ as
\eqnn\infsumno
$$\eqalignno{
\sum_{n_1=-\infty}^\infty {f_{n_1,0}}^4 \ln f_{n_1,0} 
=& {d \over dx} \sum_{n_1=-\infty}^\infty
\left(\sqrt{n_1^2}\right)^x\biggr|_{x=4} \cr
=& {d \over dx} \left(
-{\sin{\pi x \over 2} \over \pi^{{3 \over 2} + 4}}
\Gamma\left(1 + {x \over 2}\right)
\Gamma\left({1+x \over 2}\right) 
\sum_{k=-\infty \atop k\neq 0}^\infty \left(\sqrt{k^2}\right)^{-x-1}\right)\Biggr|_{x=4} \cr
=& -{1 \over 2 \pi^{9 \over 2}}\Gamma(3)\Gamma\left({5 \over 2}\right)
\sum_{k=-\infty \atop k\neq 0}^\infty {1 \over \left(\sqrt{k^2}\right)^5} \cr
=& -{1 \over \pi^{9 \over 2}}\Gamma(3)\Gamma\left({5 \over 2}\right)\zeta(5). &\infsumno
}$$
We can also show in the same way as for \infsumno\ 
that the third term in \vacEzIIi\ becomes
\eqnn\infsumon
$$\eqalignno{
\sum_{n_2=-\infty}^\infty {f_{0,n_2}}^4 \ln f_{0,n_2} 
=& {d \over dx} \sum_{n_2=-\infty}^\infty
\left(\sqrt{n_2^2 \over \sin^2\theta}\right)^x\Biggr|_{x=4} \cr
=& {d \over dx} \left(
{1 \over (\sin \theta)^x} \sum_{n_2=-\infty}^\infty
\left(\sqrt{n_2^2}\right)^x\right) \Biggr|_{x=4} \cr
=& -{1 \over \pi^{9 \over 2} \sin^4\theta}\Gamma(3)\Gamma\left({5 \over 2}\right)\zeta(5). &\infsumon
}$$
Substituting \infsumnn, \infsumno\ and \infsumon\ into \vacEzIIi, 
the vacuum energy with the condition of $(\Bbb{Z}_2)^2$ symmetry 
is written down as
\eqnn\vacEzIIii
$$\eqalignno{
E_{\Bbb{Z}_2} =& -{\Gamma(3) \over 256 \pi^7 R^4} 
\left[\Gamma(3) \sin\theta \sum_{k,l=-\infty \atop (k,l)\neq(0,0)}
{1 \over (k^2 + l^2\sin^2\theta)^3}
+ 2\sqrt{\pi}\ \Gamma\left({5 \over 2}\right)\zeta(5)
\left(1 + {1 \over \sin^4\theta}\right)\right] \cr
=& -{\Gamma(3) \over 256 \pi^7 R^4} \Biggl[
4\Gamma(3) \sin\theta \sum_{k,l=1}^\infty {1 \over (k^2 + l^2 \sin^2\theta)^3} \cr
&+ 2\Gamma(3)\zeta(6)\left(\sin\theta + {1 \over \sin^5\theta}\right)
+ 2\sqrt{\pi}\ \Gamma\left({5 \over 2}\right)\zeta(5)
\left(1 + {1 \over \sin^4\theta}\right)\Biggr]. &\vacEzIIii 
}$$
\bigskip
\vbox{\centerline{\epsfbox{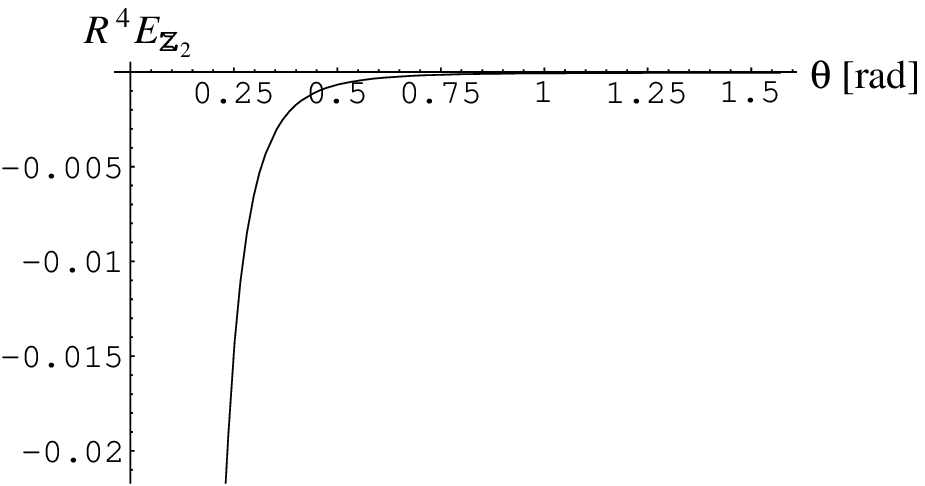}}}\nobreak
\medskip\nobreak
\centerline{\fig\figvaceneZii{The $\theta$ dependence of the vacuum energy 
with $(\Bbb{Z}_2)^2$ symmetry. The values are negative for all $\theta$.}: 
The $\theta$ dependence of the vacuum energy 
with $(\Bbb{Z}_2)^2$ symmetry.}
\nobreak
\vskip -1mm
\centerline{
The values are negative for all $\theta$.}
\bigskip\noindent
We can calculate the coefficient of $R^{-4}$ in \vacEzIIii\ numerically and 
the result is depicted in \figvaceneZii. 
The vacuum energy $E_{\Bbb{Z}_2}$ is negative for all $\theta$.

\subsec{with Dirichlet condition}\subseclab\withDCond

We require the Dirichlet condition on the boundary of 
the fundamental region of $T^2$, that is, 
on the two lines defined by $y_1 \tan\theta - y_2 = 0$ and $y_2=0$. 
In other words, the wave function $\psi(X_1,X_2)$ should vanish 
on the lines 
$X_1=0$ and $X_2=0$. So the eigen functions of \newmassope\ are proportional to
the products of sine functions, 
$$
\sin (n_1 X_1) \sin (n_2 X_2) 
= \sin n_1\left({y_1 \over R_1} - {y_2 \over R_1\tan \theta}\right) 
\sin n_2 \left({y_2 \over R_2 \sin\theta}\right), 
\quad n_1,n_2 \in \Bbb{N}. 
$$
The eigen values, that is, the masses of Kaluza-Klein modes become
\eqn\kkmassDir{
M = {1 \over R}f_{n_1,n_2}(\theta), \quad n_1,n_2 \in \Bbb{N}. 
}
By the use of \infsumnn, \infsumno\ and \infsumon, 
we obtain the vacuum energy,
 
\eqnn\vacEDir
$$\eqalignno{
E_{\rm Dirichlet} =& -{1 \over 64\pi^2} \sum_{n_1,n_2=1}^\infty 
M^4\left[\ln {\lambda^2 \over M^2} + {3 \over 2} \right] \cr
=& {1 \over 32\pi^2R^4}
\left[{1 \over 2}\sum_{n_1=-\infty}^\infty 
\left(1- \delta_{n_1,0}\right)\right]
\left[{1 \over 2}\sum_{n_2=-\infty}^\infty
\left(1 - \delta_{n_2,0}\right)\right]
{f_{n_1,n_2}}^4 \ln f_{n_1,n_2} \cr
=& {1 \over 128\pi^2R^4}\left[\sum_{n_1,n_2=-\infty}^\infty 
{f_{n_1,n_2}}^4\ln f_{n_1,n_2}
- \sum_{n_1=-\infty}^\infty {f_{n_1,0}}^4 \ln f_{n_1,0}
- \sum_{n_2=-\infty}^\infty {f_{0,n_2}}^4 \ln f_{0,n_2}\right] \cr
=& -{\Gamma(3) \over 256 \pi^7 R^4} \Biggl[
4\Gamma(3)\sin\theta \sum_{n_1,n_2=1}^\infty {1 \over (n_1^2 + n_2^2 \sin^2\theta)^3} \cr
& + 2\Gamma(3)\zeta(6)\left(\sin\theta + {1 \over \sin^5\theta}\right)
- 2\sqrt{\pi}\ \Gamma\left({5 \over 2}\right)\zeta(5)
\left(1 + {1 \over \sin^4\theta}\right)\Biggr]. &\vacEDir
}$$

\bigskip
\vbox{\centerline{\epsfbox{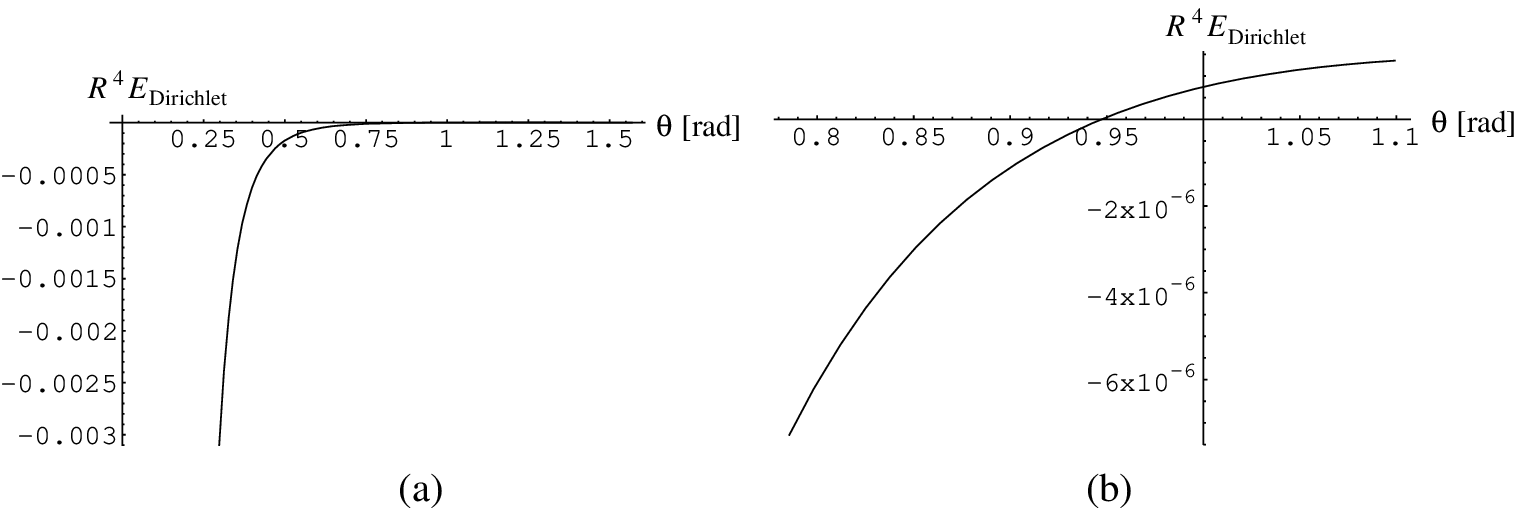}}}\nobreak
\medskip\nobreak
\centerline{\fig\figvaceneDir{the $\Theta$ dependence of the vacuum energy 
with Dirichlet condition. The values turn from negative to positive at 
$\theta=\theta_{\rm cr}$.}: The $\theta$ dependence of the vacuum energy 
with Dirichlet condition. The values turn }
\nobreak
\vskip -1mm
\centerline{
from negative to positive at 
$\theta=\theta_{\rm cr}$ as shown in the magnified plot (b).}
\bigskip\noindent

The results of the numerical calculations of $R^4E_{\rm Dirichlet}$ 
are represented in \figvaceneDir.
In this case it is significant that the positive vacuum energy appears. 
The vacuum energy $E_{\rm Dirichlet}$ is equal to zero at 
$\theta = \theta_{\rm cr}$ (\figvaceneDir (b)).
For $\theta < \theta_{\rm cr}$ the vacuum energy is negative, 
while for $\theta > \theta_{\rm cr}$ it is positive. 
The critical angle $\theta_{\rm cr}$ is approximately equal 
to 0.947243 [rad] which is calculated numerically.

The positive vacuum energy is favourable, because the cosmological constant 
is positive. 
From this point of view, we will give some comments in the next section.

\newsec{Comments on Generic Size of Extra Dimensions}

The extra dimensions have been introduced in \refs{\ADD,\AADD} 
in order to solve the hierarchy problem . 
When we consider a $4+d$ dimensional space-time, that is, 
a $d$ dimensional extra space, the relation between 
the Plank scale $M_{\rm pl}$ of our four dimensional world 
and the one $M_{{\rm pl}(4+d)}$ of the $4+d$ dimensional space-time 
is described by
$$
{M_{\rm pl}}^2 \sim {M_{{\rm pl}(4+d)}}^{2+d}R^d,
$$
where $R$ denotes the size of the extra dimensions. 
When we assume that $M_{{\rm pl}(4+d)} \sim m_{\rm EW} \sim 1$[TeV], 
we obtain the size for the extra space as 
$$
R \sim 10^{{30 \over d}-17} {\rm [cm]}.
$$
For $d=2$, the size $R$ becomes a sub-millimeter size, $10^{-2}$[cm].

On the other hand, we have not had a complete answer for the cosmological 
constant problem yet. But a lot of features have been made clear by 
some observations, for a recent example, the WMAP observation. 
The results of the WMAP\refs{\WMAPi,\WMAPii} provide us the following data;
\eqna\datawmap
$$\eqalignno{
h &= 0.71 \quad \hbox{(Hubble constant)}, &\datawmap{a}\cr
\Omega_{\rm tot} &= 1.02\pm 0.02 \quad \hbox{(total density)}, &\datawmap{b}\cr
\Omega_\Lambda &= 0.73\pm 0.04 \quad \hbox{(dark energy density)}. &\datawmap{c}
}$$
From \datawmap{a}, the Hubble constant is 
\eqn\hubblewmap{
H_0 = h \times 100 [{\rm km}\cdot{\rm sec}^{-1}\cdot{\rm Mpc}^{-1}] 
= 0.7675\times 10^{-28}[{\rm cm}^{-1}].
}
In terms of \hubblewmap, we obtain the critical density, 
\eqn\cdwmap{
\rho_{\rm c} = {3H_0^2 \over 8\pi G} 
= 5.313 \times 10^3 [{\rm eV}\cdot{\rm cm}^{-3}].
}
Since we consider that the dark energy is almost the same 
as the cosmological constant $\Lambda$, 
it is calculated from \datawmap{b}, \datawmap{c} and \cdwmap\ as
\eqn\ccwmap{
\Lambda = \rho_{\rm c} \Omega_{\rm tot} \Omega_\Lambda
= 3.956\times 10^3 [{\rm eV}\cdot{\rm cm}^{-3}].
}

Now we assume that the cosmological constant is identified with 
the vacuum energy to which the Kaluza-Klein modes from the compactification 
of extra dimensions contribute. 
The vacuum energy is proportional to $R^{-4}$. 
If the size of the extra space $R$ is a submillimeter size, 
$R^{-4}$ is in the order of 
$10^8 [{\rm cm}^{-4}] \sim 10^4 [{\rm eV}\cdot{\rm cm}^{-3}]$. 
Since it is close to the cosmological constant \ccwmap, 
that identification is reasonable.
The relationship between the vacuum energy, in other words, 
the Casimir energy, and the cosmological constant has been discussed in 
the various papers \refs{\MS,\Ro\Diii\PePo\Ga\Mi\Ch\Me\Gu\Mii\BLS{--}\El}.
We believe from recent observations that the cosmological constant 
is positive. 
On account of that identification, the vacuum energy also has to be positive. 
Fortunately we were able to realize a positive vacuum energy by the use of 
the extra dimensions with the Dirichlet condition in the section \withDCond. 
\bigskip
\vbox{\centerline{\epsfbox{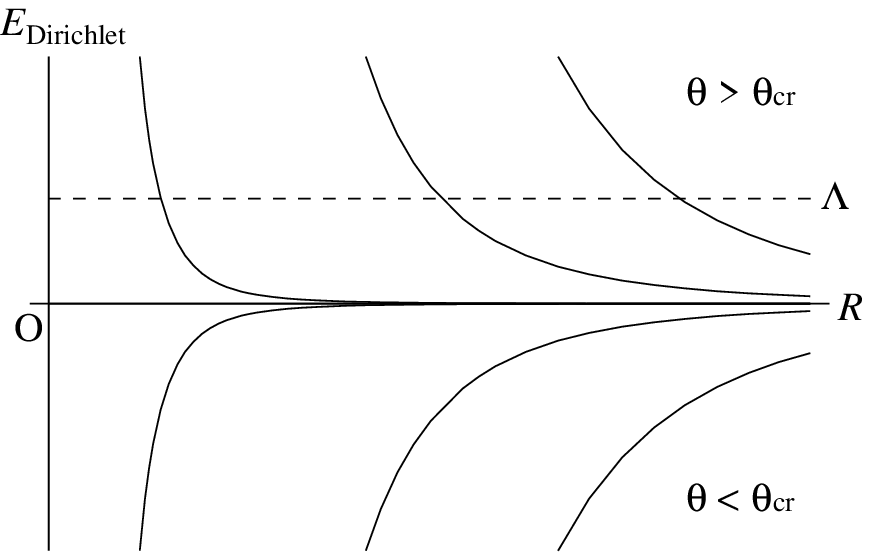}}}\nobreak
\medskip\nobreak
\centerline{\fig\figcsmcnst{}: The $R$ dependence of the vacuum energy. 
If $|\theta - \theta_{\rm cr}|$ increases, the corresponding line }
\nobreak
\vskip -1mm
\centerline{for the graph of $(R,E_{\rm Dirichlet} )$ moves away from 
the vertical axis.}

\bigskip\noindent
Since the vacuum energy $E_{\rm Dirichlet}$ depends on $R^{-4}$ from \vacEDir, 
the behaviour of $E_{\rm Dirichlet}$ for $R$ is represented by \figcsmcnst.

If the values of the cosmological constant $\Lambda$ and 
the moduli parameter $\theta$ are given, 
the size of extra dimensions $R$ is determined. 
For example, let us consider the case of $\theta = \pi/2$. 
This is the situation in which the torus $T^2$ has no more shape moduli 
and is essentially equal to the product of two spheres, $(S^1)^{\otimes 2}$, 
with the same radius $R$. The vacuum energy \vacEDir\ then becomes
$E_{\rm Dirichlet} = 1.1769 \times 10^{-6} / R^4$.
In terms of $\hbar c = 1.973269 \times 10^{-5} [{\rm eV}\cdot{\rm cm}]$, 
we obtain 
$$
E_{\rm Dirichlet} [{\rm eV}\cdot{\rm cm}^{-3}] 
= {2.3223 \times 10^{-11} \over (R [{\rm cm}])^4}.
$$
When we identify this vacuum energy with the cosmological constant \ccwmap, 
the size of the extra dimensions $R$ becomes
$2.768 \times 10^{-4}$[cm]. 
Though this scale is different by $10^{-2}$ order from the one derived 
from the hierarchy problem \refs{\ADD,\AADD}, 
it is not so crucial an inconsistency. 

Note that there is no zero mode under the Dirichlet condition.
This is a serious problem, if the particle propagating in the extra space is 
a graviton. But it is not a problem for a scalar particle, 
because, in fact,  there is no such a massless scalar particle 
in our four dimensional space-time. 

\newsec{Discussions and Conclusions}

We considered the torus $T^2$ as a two dimensional extra space. 
We introduced the shape moduli of $T^2$ and analysed the dependence of 
the vacuum energy, or the Casimir energy, 
on the shape moduli parameter, 
to which all the Kaluza-Klein modes 
derived from the $T^2$ compactification contributed. 
Without a condition on the covering space of $T^2$ 
except for bi-periodic condition, we can find eigen functions 
for the standard mass operator. 
But, if we require the $(\Bbb{Z}_2)^2$ or the Dirichlet conditions, 
there is no eigen function for the mass operator. 
So we proposed the new mass operator \newmassope\ 
together with the corresponding eigen functions 
and 
considered the vacuum energy for this operator.

In terms of the mathematical tools of 
the Fourier transformation and the generalized functions, 
we regularized the divergence of the infinite summations 
of Kaluza-Klein modes. 
For a specific case of $d=1$, 
our regularization reduces to the zeta-function regularization, 
where higher mode is suppressed and infinity arising 
in infinite volume limit is subtracted. 
This regularization provides us the same results 
as are given by the cuttoff method \AC. 
Our method being generalized to multi-dimensional problems 
is a valid and unique one 
to the extent that the zeta-function regularization is a valid method.

We then made it possible to calculate numerically the vacuum energy 
including the contributions of all the Kaluza-Klein modes. 
In general the vacuum energy has a negative value. 
But we have found that, if we impose the Dirichlet conditions, 
the value of the vacuum energy turns out to be positive 
lifting from a negative territory, 
as the shape moduli parameter changes crossing the critical value.

We assumed that the vacuum energy was identified 
with the cosmological constant.
Since it is confirmed by the observations, for example, the WMAP observation, 
that the cosmological constant is positive, 
we compared the WMAP's observational value of 
the cosmological constant with 
the vacuum energy which we calculated under the Dirichlet condition. 
From this comparison we showed that the size of extra dimensions is 
of order $10^{-4}$[cm]. 
The value of $R = 2.768 \times 10^{-4}$[cm] presented 
in the section 4 is the maximal one that we can realize in the model 
considered in this paper. 
On the other hand, from the hierarchy problem, 
the size of extra space for the two extra dimensions is $10^{-2}$[cm]. 
Though these two sizes do not accurately agree with each other, 
they are sufficiently consistent. 

Since the cosmological constant may consist not only of the vacuum energy 
but also of some other elements, 
it is significant that we showed that the two sizes of extra dimensions 
derived from the hierarchy problem and from the cosmological constant 
problem are similar. 

It is possible to consider various models with an extra space. 
So one of future problems is for us to find the model in which 
the vacuum energy has a positive value of right order of magnitude. 
Since we calculated the vacuum energy only in a semi-classical limit 
in this paper, 
it is also a future problem to consider possible quantum loop effects 
and also some possible contributions coming from quantum fluctuations of 
relevant metric components.

\bigbreak\bigskip\bigskip
\centerline{{\bf Acknowledgements}}\nobreak

S.S. is grateful to N. Maru and M. Tachibana for useful discussions. 
S.M. was supported in part by Grant-in-Aid for Scientific 
Research on Priority Areas 763 
``Dynamics of Strings and Fields,'' while 
S.S. by Grant-in-Aid for the 21st Century COE 
``Center for Diversity and Universality in Physics,''
from the Ministry of Education, 
Culture, Sports, Science and Technology, Japan. 

\listrefs
\bye